\documentclass[%
 reprint,
superscriptaddress,
amsmath,amssymb,
aps,
prb,
]{revtex4-2}

\usepackage{graphicx}
\usepackage{dcolumn}
\usepackage{bm}
\usepackage{braket}
\usepackage{float} 
\usepackage{amsmath}
\usepackage{amsfonts}   
\usepackage{amssymb}    
\usepackage{comment}
\usepackage[table]{xcolor} 
\usepackage{colortbl}      

\usepackage{array}
\newcolumntype{P}[1]{>{\centering\arraybackslash}p{#1}}

\begin{document}

\preprint{APS/123-QED}

\title{Anomalous Order in $R$\textsubscript{3}Ni\textsubscript{30}B\textsubscript{10} ($R$ = La, Ce)}

\author{Maximilien F. Debbas}
\affiliation{Department of Nuclear Science and Engineering, Massachusetts Institute of Technology, Cambridge, MA 02139, USA}

\author{Takehito Suzuki}
\affiliation{Department of Physics, Toho University, Funabashi, Japan}
 
\author{Joseph G. Checkelsky}%
\affiliation{Department of Physics, Massachusetts Institute of Technology, Cambridge, MA 02139, USA}%

\date{\today}

\begin{abstract}

We report the synthesis of single crystals of the $R$\textsubscript{3}Ni\textsubscript{30}B\textsubscript{10} ($R$ = La, Ce) system which realizes the tetragonal $P4/nmm$ (No. 129) space group. We performed single crystal, transmission X-ray diffraction measurements to determine the crystal structure. Additionally, we characterized the samples through magnetization, resistivity, and heat capacity measurements. The $R=$ Ce system exhibits mixed-valence states and we observe maxima in the heat capacity at $T_1^* = 6 \, \text{K}$ and $T_2^* = 2 \, \text{K}$ (both absent for $R=$ La) with no corresponding features in the resistivity or magnetization. We discuss the potential roles of multipolar and short-ranged/partial order in connection to the $T_1^*$ anomalous order.

\end{abstract}

\maketitle


\section{\label{sec:Introduction} Introduction}

Rare earth systems are known to realize a multitude of different phases arising from their localized \textit{f}-electron states provided by lanthanide or actinide elements in the crystal. An important subset of these materials is that able to realize long-range multipolar order across a lattice of localized \textit{f}-electron states. The most well known of these systems are the praseodymium-based quadrupolar order materials which take advantage of the integer-quantized $J=4$ ground state of the Pr\textsuperscript{3+} ion. Kramers' theorem does not apply to this angular momentum state which allows for a crystal electric field (CEF) to split the free-space $J=4$ ground state into potentially nonmagnetic multiplets realizing a leading-order quadrupole moment. Notable examples of praseodymium systems realizing long-range quadrupolar order include PrIr\textsubscript{2}Zn\textsubscript{20} \cite{PhysRevLett.106.177001}, PrRh\textsubscript{2}Zn\textsubscript{20} \cite{PhysRevB.86.184426}, PrV\textsubscript{2}Al\textsubscript{20} \cite{sakai2011kondo}, PrTi\textsubscript{2}Al\textsubscript{20} \cite{sakai2011kondo}, and PrPb\textsubscript{3} \cite{PrPb3_2001}. Some cerium-based systems are also known to host quadrupolar order. As the Ce\textsuperscript{3+} ion ground state realizes the half-integer $J=5/2$ angular momentum, Kramers' theorem requires that all CEF split energy levels be magnetic. Notable materials such as CeB\textsubscript{6} \cite{19841809} and Ce\textsubscript{3}Pd\textsubscript{20}Ge\textsubscript{6} \cite{SUZUKI1999334} will then realize multiple phases with distinct quadrupolar and magnetic ordering temperatures. Multipolar interactions (sometimes involving higher-order multipoles) can yield rich phase diagrams in systems such as Ce$_{1-x}$La$_x$B$_6$ \cite{PhysRevLett.95.117206,doi:10.1143/JPSJ.53.3967}, YbRu\textsubscript{2}Ge\textsubscript{2} \cite{PhysRevB.73.020407,PhysRevB.77.045105}, URu\textsubscript{2}Si\textsubscript{2} \cite{ohkawa1999quadrupole,AMITSUKA2007214,BOURDAROT2005986,doi:10.7566/JPSJ.84.024717}, U\textsubscript{0.75}Np\textsubscript{0.25}O\textsubscript{2} \cite{PhysRevB.70.214402}, and NpO\textsubscript{2} \cite{caciuffo2003multipolar,PhysRevLett.89.187202,PhysRevLett.97.207203}.

Some rare earth compounds may also exhibit mixed-valance behavior and move the system away from the lattice picture of localized, \textit{f}-electron moments; such mixed-valence behavior is especially prevalent in cerium and ytterbium based systems \cite{PhysRevB.43.10906, KISHIMOTO2003308, MOROZKIN2016290}. In mixed-valance cerium systems, the cerium \textit{f}-state realizes a valence between that of nonmagnetic Ce\textsuperscript{4+} ($4f^0$) and magnetic Ce\textsuperscript{3+} ($4f^1$) such that the \textit{f}-electrons are partially delocalized \cite{PhysRevB.43.10906, LEGVOLD1980183}. Some mixed-valence cerium systems are known to realize long-ranged magnetic order despite the partially delocalized $f$-states, \textit{viz.} CeRuSn \cite{PhysRevB.87.094421, fikavcek2013nature} and Ce\textsubscript{3}Rh\textsubscript{4}Sn\textsubscript{7} \cite{OPLETAL2022166941}. In other mixed-valence compounds such as CeNiSi\textsubscript{2}, fluctuations in the partially delocalized \textit{f}-states can lead to short-ranged correlations at low temperature which manifest themselves as broad maxima in the heat capacity corresponding to low losses of entropy \cite{PhysRevB.43.10906}. Short-ranged quadrupolar fluctuations are also known to yield similar behavior in the doped Y$_{1-x}$Pr$_{x}$Ir$_{2}$Zn$_{20}$ system \cite{PhysRevLett.121.077206}. 

New material systems in this class provide an opportunity to not only study the properties of $f$-electron states in unconventional settings, but also provide potential insight into their design.  Herein, we report a new single crystal system Ce\textsubscript{3}Ni\textsubscript{30}B\textsubscript{10} hosting mixed-valence cerium. Through combined thermodynamic and transport studies, we find evidence for anomalous order manifesting a small entropy response in heat capacity but not other observables. We discuss possible mechanisms for this and how tuning this system may provide an opportunity to study connections between the disparate phenomena of metallic systems containing $f$-electrons.

\section{\label{sec:Growth Method} Growth Method}

Single crystals of the $R$\textsubscript{3}Ni\textsubscript{30}B\textsubscript{10} system ($R$ = La, Ce) were synthesized through a nickel-boron flux which takes advantage of the 1018\textsuperscript{o}C eutectic point occurring at a nickel to boron ratio of 11:9. The samples were prepared by employing a $R$:Ni:B ratio of 4.1:11:9 loaded into an alumina crucible sealed in a quartz tube under high vacuum. The materials used were high-purity, lump cerium and lanthanum ingots (AMES), 99.9999\% purity, -4 mesh boron powder (Thermo Scientific), and 99.996\% purity, -120 mesh nickel powder (Thermo Scientific). The cerium and lanthanum ingots were, respectively, clipped into chunks and filed into shavings under argon in a glovebox.

The elements for the $R=$ Ce growth were melted together by first heating to 1100\textsuperscript{o}C over 12 hours and holding at that temperature for four days. After this, the material was cooled to 850\textsuperscript{o}C over a week and then finally cooled to room temperature over 12 hours. The $R=$ La materials were initially heated to 1150\textsuperscript{o}C over 12 hours and then held at that temperature for two days after which they were cooled to 900\textsuperscript{o}C over a week and then finally cooled to room temperature over another 12 hours. 

\begin{figure}[H]
	\centering 
	\includegraphics[width=1\linewidth]{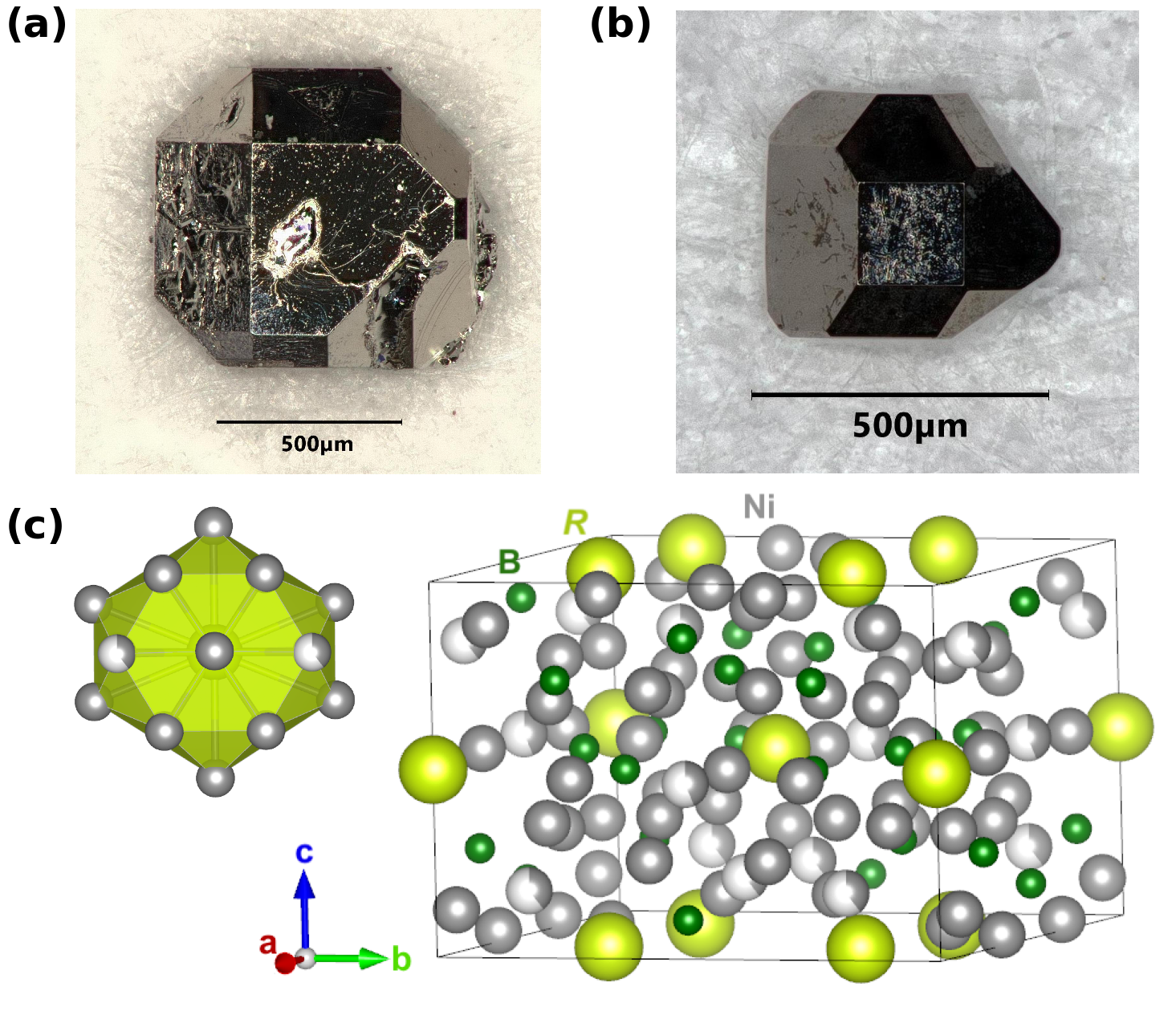}
	\caption{(a) La\textsubscript{3}Ni\textsubscript{30}B\textsubscript{10} single crystal. (b) Ce\textsubscript{3}Ni\textsubscript{30}B\textsubscript{10} single crystal. (c) $R$\textsubscript{3}Ni\textsubscript{30}B\textsubscript{10} ($R$ = La, Ce) crystal structure showing the lanthanide in yellow, the nickel in grey, and the boron in green; the grey/white sites indicate nickel positions of partial occupancy. The inset (left) shows the coordination polyhedron for each lanthanide site with the primary rotational symmetry axis perpendicular to the page. Schematics were drawn using VESTA \cite{VESTA}.}
	\label{fig:1}
\end{figure}

Both growths resulted in a large number of small ($<$1 mm length) crystals and a few large ($>$1 mm length) crystals embedded in a matrix of shiny, grey material in the alumina crucible. The crystals were mechanically separated out of the this matrix. Figure \ref{fig:1}(a) shows a La\textsubscript{3}Ni\textsubscript{30}B\textsubscript{10} crystal and Fig. \ref{fig:1}(b) shows a Ce\textsubscript{3}Ni\textsubscript{30}B\textsubscript{10} crystal.

\section{\label{sec:Crystal Structure} Crystal Structure}

The $R$\textsubscript{3}Ni\textsubscript{30}B\textsubscript{10} ($R$ = La, Ce) system realizes a structure in the tetragonal $P4/nmm$ (No. 129) space group to our knowledge not previously reported within the $R$-Ni-B phase diagram. Transmission geometry X-ray diffraction was performed on single crystals at $100$ K with Mo-$K_\alpha$ radiation; a full-matrix, least-squares fit on $F^2$ was performed using SHELXL to determine and refine the structure. Crystallographic data is provided in appendix \ref{sec: X-ray Diffraction} for both the $R$ = La and $R$ = Ce systems.

The refined crystal structure includes eight nickel Wyckoff positions; two of these positions exhibit a partial occupancy of approximately $40 \%$ nickel. Were all the sites fully occupied, the system would be isostructural to the Nd\textsubscript{3}Ni\textsubscript{29}Si\textsubscript{4}B\textsubscript{10} system \cite{ZHANG1998239}. The stoichiometric structure with full site occupancy would likely be difficult to synthesize using the method described here as adding nickel to the growth would prevent the use of the 1018\textsuperscript{o}C nickel-boron eutectic point.

Figure \ref{fig:1}(c) shows the $R$\textsubscript{3}Ni\textsubscript{30}B\textsubscript{10} unit cell. The system adopts a structure of lanthanide sites each surrounded by a coordination polyhedron of nickel sites with a network of boron interspersed between them. Each lanthanide atom coordinates with 20 nickel sites to form a distorted crystal field (shown in the inset of Fig. \ref{fig:1}(c)). Were it undistorted, this coordination polyhedron would realize the point group $D_{6h}$; the distortion lowers the symmetry group to $D_2$ (see appendix \ref{sec: Local Crystal Field} for additional details). Note that the $P4/nmm$ space group is globally centrosymmetric and the undistorted coordination polyhedron locally preserves inversion symmetry around each lanthanide site.

\section{\label{sec:Transport, Thermodynamic, and Spectroscopic Measurements} Transport, Thermodynamic, and Spectroscopic Measurements}

Herein, we present results derived from resistivity, magnetization, heat capacity, and X-ray photoelectron spectroscopy (XPS) measurements of three Ce\textsubscript{3}Ni\textsubscript{30}B\textsubscript{10} crystals (\textit{C1}, \textit{C2}, and \textit{C3}) which were all grown in the same synthesis batch. Magnetization and heat capacity measurements on the La\textsubscript{3}Ni\textsubscript{30}B\textsubscript{10} system were also performed on a single crystal (\textit{L1}) and serve as a reference to which the cerium $f$-electron physics may be compared. The crystals used for the measurements presented herein were not oriented prior to measurement (we did not observe a strong dependence on orientation, though a more detailed study is the subject of future work).

\subsection{Resistivity}
\label{Resistivity}

Electrical transport measurements were performed by a conventional five-probe method in a Quantum Design Physical Property Measurement System (PPMS). Ce\textsubscript{3}Ni\textsubscript{30}B\textsubscript{10} crystal \textit{C2} was sanded down to a thickness of $57 \, \mu\text{m}$ with a commercial lapping jig and affixed to a sapphire substrate with GE varnish. Gold wire contacts were affixed to the crystal using silver paint. 

\begin{figure}[H]
	\centering 
	\includegraphics[width=1\linewidth]{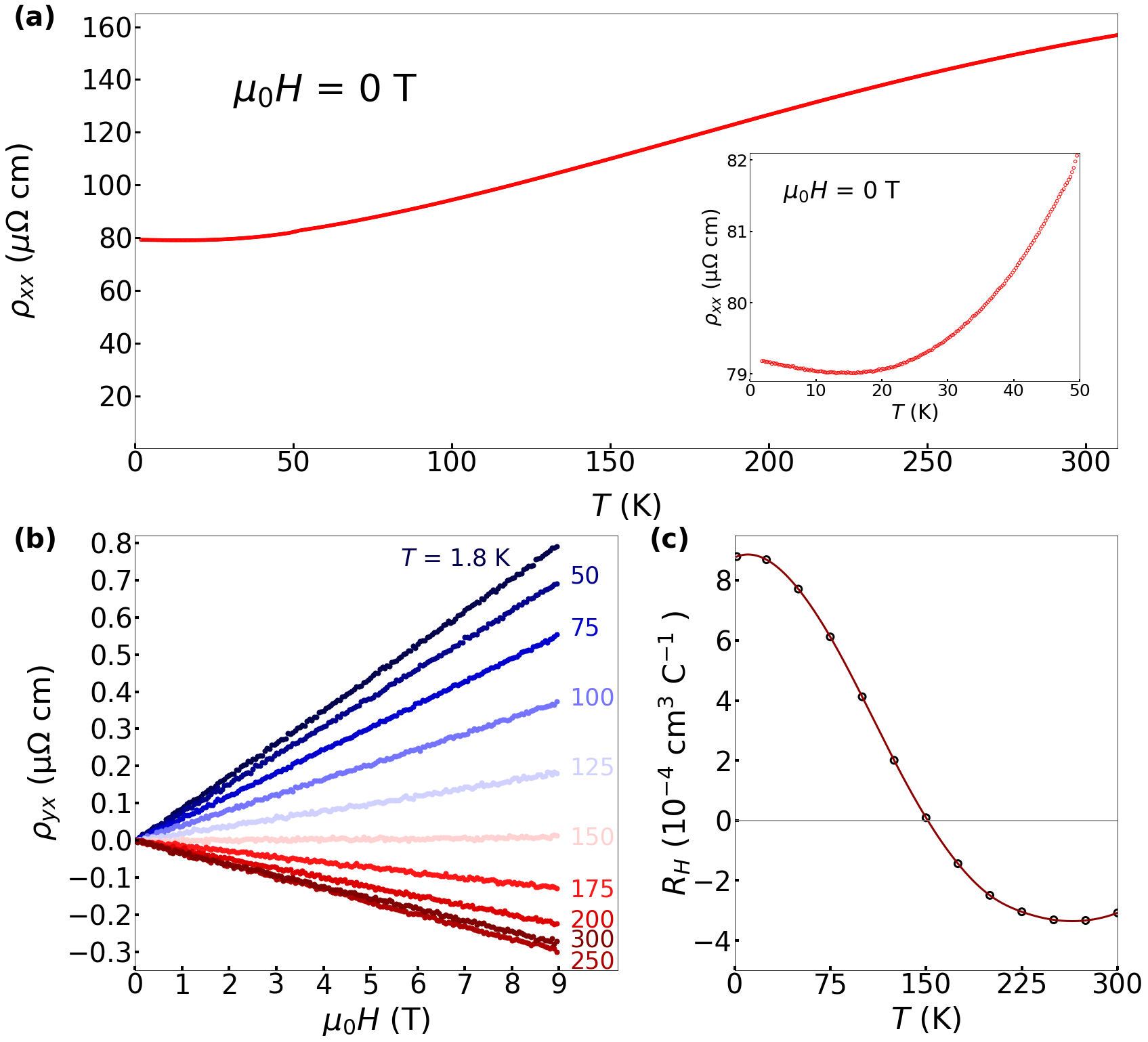}
	\caption{(a) Zero-field resistivity of Ce\textsubscript{3}Ni\textsubscript{30}B\textsubscript{10} crystal \textit{C2} (the inset shows the resistivity below $50 \, \text{K}$). (b) Hall resistivity of crystal \textit{C2} plotted as a function of applied magnetic field for several temperatures. (c) Hall coefficient of the same crystal plotted as a function of temperature. The smooth curve is a guide to the eye.}
	\label{fig:2}
\end{figure}

Figure \ref{fig:2}(a) shows the zero-field longitudinal resistivity of Ce\textsubscript{3}Ni\textsubscript{30}B\textsubscript{10} as a function of temperature below $300 \, \text{K}$, and the inset shows the resistivity below $50 \, \text{K}$. The resistivity exhibits a broad minimum at around $20 \, \text{K}$ as well as an inflection point around $175 \, \text{K}$. Due to this inflection point, the resistivity is not well fit by the Bloch-Gr{\"u}neisen model at temperatures above the broad minimum, though the overall response is metallic.

Figure \ref{fig:2}(b) shows Hall resistivity data as a function of applied magnetic field at various temperatures and Fig. \ref{fig:2}(c) shows the associated Hall coefficient. This measurement indicates a primarily electron-like response at high temperature and a hole-like response at low temperature. The crossover temperature between these two regimes occurs near $150 \, \text{K}$. Using a single band approximation, we find carrier densities of $n_{e} = 2.0 \times 10^{22} \, \text{cm}^{-3}$ at $300 \, \text{K}$, and $n_{h} = 7.1 \times 10^{21} \, \text{cm}^{-3}$ at $1.8 \, \text{K}$.

\subsection{Magnetization}
\label{Magnetization}

Magnetic susceptibility $\chi$ and magnetization $M$ were measured using SQUID magnetometry in a Quantum Design Magnetic Property Measurement System (MPMS3). Single crystals were affixed to a quartz rod using GE varnish and DC magnetization measurements were performed using Vibrating Sample Magnetometer (VSM) mode.

\begin{figure}[H]
	\centering 
	\includegraphics[width=1\linewidth]{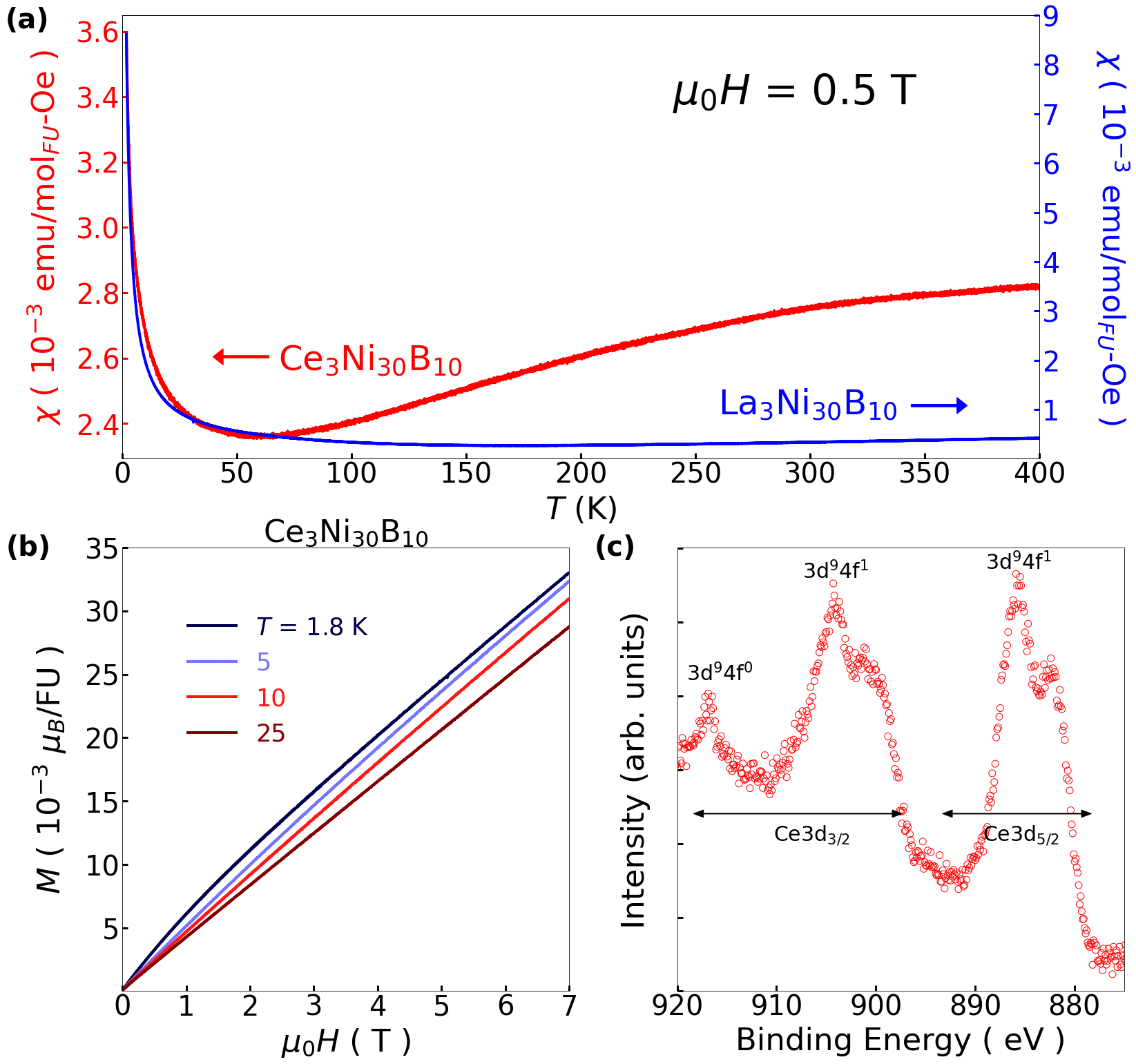}
	\caption{(a) Magnetic susceptibility data for Ce\textsubscript{3}Ni\textsubscript{30}B\textsubscript{10} crystal \textit{C2} and La\textsubscript{3}Ni\textsubscript{30}B\textsubscript{10} crystal \textit{L1} measured in $0.5 \, \text{T}$ of applied magnetic field. (b) Magnetization ($\mu_B$ per formula unit) plotted as a function of applied field for Ce\textsubscript{3}Ni\textsubscript{30}B\textsubscript{10} crystal \textit{C2}. (c) Ce 3$d$ XPS spectrum for Ce\textsubscript{3}Ni\textsubscript{30}B\textsubscript{10} crystal \textit{C3} showing the 3$d_{5/2,3/2}$ doublet. The most prominent final states dependent on $f$-electron filling are also labeled.}
	\label{fig:3}
\end{figure}

Figure \ref{fig:3}(a) shows the magnetic susceptibility for both Ce\textsubscript{3}Ni\textsubscript{30}B\textsubscript{10} and La\textsubscript{3}Ni\textsubscript{30}B\textsubscript{10} measured in an applied magnetic field of $0.5 \, \text{T}$. Both systems lack any sharp features; they exhibit paramagnetic behavior down to base temperature. This indicates that neither the cerium nor the nickel realizes any prominent magnetic ordering in either compound. Magnetic torque (data not shown here) measured as a function of both temperature ($1.8 \, \text{K}$ - $300 \, \text{K}$) and field (up to $14 \, \text{T}$) also does not show any signatures of a magnetic transition.

Neither system exhibits simple Curie-Weiss behavior of the magnetic susceptibility over a broad temperature range. The inverse susceptibilities of Ce\textsubscript{3}Ni\textsubscript{30}B\textsubscript{10} and La\textsubscript{3}Ni\textsubscript{30}B\textsubscript{10} exhibit broad maxima around $50 \, \text{K}$ and $150 \, \text{K}$ respectively (see appendix \ref{sec: Curie-Weiss Fit of Magnetic Susceptibility} for additional details). At low temperatures, the magnetic susceptibilities can be fit to a simple Curie-Weiss model with a constant offset. Fitting the Ce\textsubscript{3}Ni\textsubscript{30}B\textsubscript{10} susceptibility below $30 \, \text{K}$ yields a Curie-Weiss temperature of $\theta_{CW} = -2.27 \, \text{K}$ and an effective moment of $0.201 \, \mu_B$ per formula unit. Fitting the La\textsubscript{3}Ni\textsubscript{30}B\textsubscript{10} susceptibility below $50 \, \text{K}$ yields $\theta_{CW} = -0.733 \, \text{K}$ and an effective moment of $0.395 \, \mu_B$ per formula unit.

Figure \ref{fig:3}(b) shows the Ce\textsubscript{3}Ni\textsubscript{30}B\textsubscript{10} magnetization (in $\mu_B$ per formula unit) plotted as a function of applied magnetic field. The magnetization appears to increase almost linearly with a slight bend at low fields and does not appear to saturate up to $7 \, \text{T}$ of applied field. This, along with the low temperature susceptibility behavior, suggests that the system possesses a small amount of localized moment (Brillouin function magnetization) from the Ce/Ni sites or potential magnetic impurities in addition to a relatively large Pauli paramagnetic response (linear magnetization) from the conduction states.

\subsection{XPS}
\label{XPS}

X-ray photoelectron spectroscopy (XPS) was performed using a commercial Thermo Scientific Nexsa system with an aluminum K$\alpha$ source. Ce\textsubscript{3}Ni\textsubscript{30}B\textsubscript{10} crystal \textit{C3} was affixed to a piece of silicon wafer using silver paint, and the surface milled with an argon ion beam prior to the measurement.

Figure \ref{fig:3}(c) shows an XPS spectrum taken in the Ce 3$d$ region for Ce\textsubscript{3}Ni\textsubscript{30}B\textsubscript{10}. The peaks corresponding to the 3$d_{3/2}$ and 3$d_{5/2}$ multiplets are clearly resolved as are some of the peaks corresponding to splitting of the multiplets due to the $f$-shell filling. The final states $3d^94f^0$ and $3d^94f^1$ are clearly resolved for the 3$d_{3/2}$ multiplet, and the $3d^94f^1$ final state is well resolved for the 3$d_{5/2}$ multiplet. The 3$d_{3/2}$ $3d^94f^2$ final state and the 3$d_{5/2}$ $3d^94f^0$ final state are expected to be weak and may account for the excess counts present around $895 \, \text{eV}$. The shoulder peaks to the right of the $3d^94f^1$ final states of each multiplet are likely due to oxides as is the case in CeNi\textsubscript{4}B \cite{tolinski2003mixed}.

\subsection{Heat Capacity}
\label{Heat Capacity}

Heat capacity was measured using a Quantum Design PPMS with the heat capacity module enabled. Single crystals were affixed to heater platform using Apiezon N grease. A background heat capacity measurement was taken of the grease prior to mounting the crystal such that it could be subtracted from the total heat capacity after the crystal was mounted.

\begin{figure}[H]
	\centering 
	\includegraphics[width=1\linewidth]{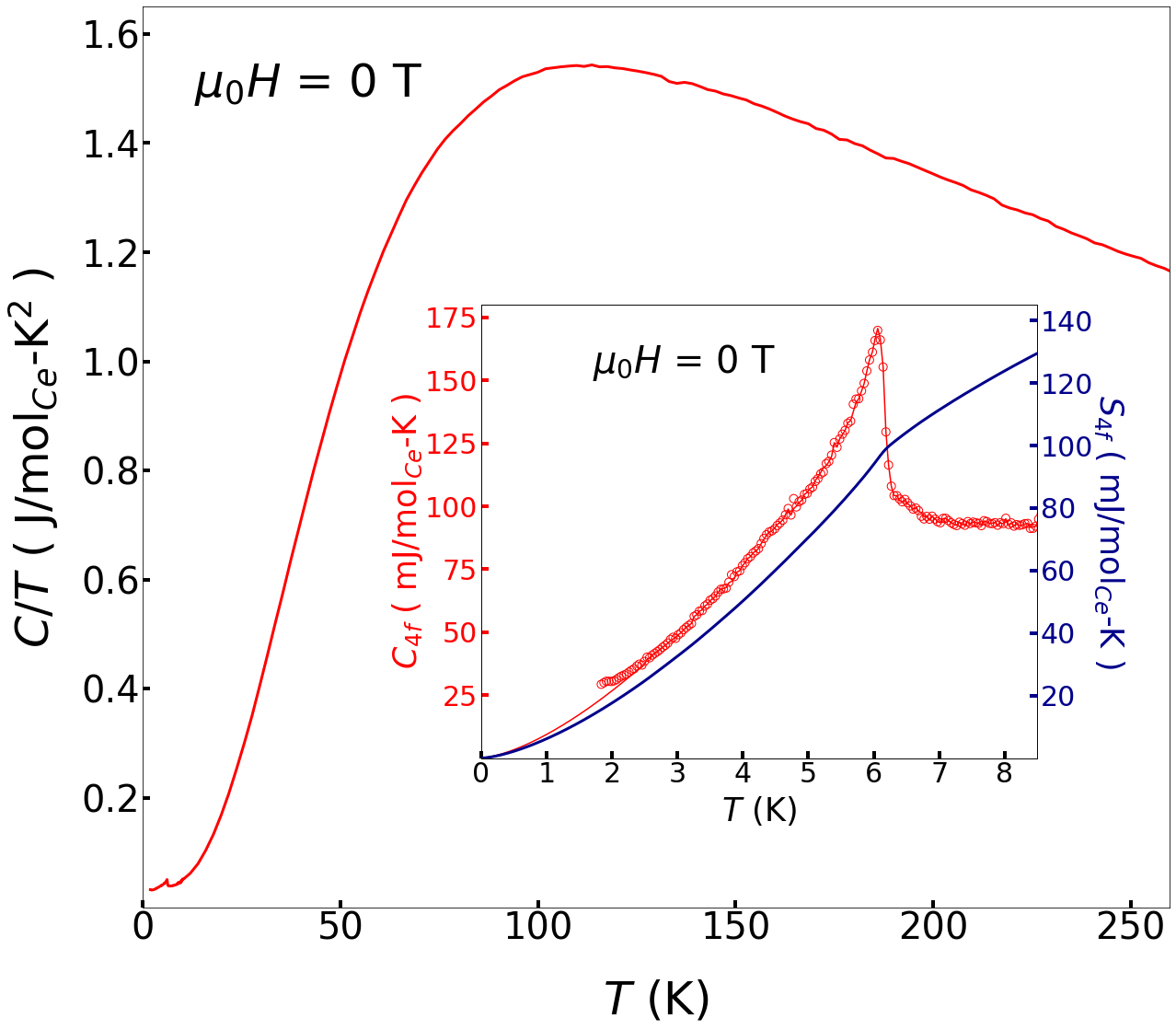}
	\caption{Ce\textsubscript{3}Ni\textsubscript{30}B\textsubscript{10} crystal \textit{C2} zero-field heat capacity divided by temperature plotted as a function of temperature. The inset shows the $4f$ contribution to heat capacity (left axis) and entropy (right axis) of the same crystal under $8 \, \text{K}$. The background heat capacity was provided by La\textsubscript{3}Ni\textsubscript{30}B\textsubscript{10} crystal \textit{L1}.}
	\label{fig:4}
\end{figure}

Figure \ref{fig:4} shows the zero-field heat capacity ($C$) divided by temperature ($T$) for Ce$_{3}$Ni$_{30}$B$_{10}$ while the inset shows the $4f$ electron contribution to the heat capacity ($C_{4f}$) as well as the calculated $4f$ entropy ($S_{4f}$) below 8 K. $C_{4f}$ was estimated by subtracting the heat capacity of La$_{3}$Ni$_{30}$B$_{10}$ from that of Ce$_{3}$Ni$_{30}$B$_{10}$. $S_{4f}$ was computed by interpolating $C_{4f}$ from base $T$ to zero using a $C \sim T^{\alpha}$ power law fit of the data between 2.5 and 5.5 K. The heat capacity exhibits a maximum at $T_1^* = 6 \, \text{K}$; at $T_1^*$, the excess heat capacity associated with the $4f$ state for crystal \textit{C2} is approximately $170 \, \text{mJ/mol\textsubscript{Ce}-K}$ and the excess entropy is approximately $95 \, \text{mJ/mol\textsubscript{Ce}-K}$).

We also note that an additional smaller peak may be present at $T_2^* = 2 \, \text{K}$ and is observed systematically in the crystals studied here. This is an important subject for future low temperature investigation. For both peaks the precise height and prominence over the background heat capacity varied, suggesting a potential interplay with disorder. Appendix \ref{sec: Ce3Ni30B10 Heat Capacity Variations} discusses heat capacity measurements for four additional crystals of the same synthesis batch. The calculated entropy at $T_1^*$ for these crystals ranged from about $95$ to $220$ mJ/mol\textsubscript{Ce}-K and were associated with power law exponents (from the $C\sim T^\alpha$ fit between $2.5 \, \text{K}$ and $5.5 \, \text{K}$) ranging from $0.7$ to $2.5$. 

Fitting the low temperature heat capacity of La\textsubscript{3}Ni\textsubscript{30}B\textsubscript{10} yields a Sommerfeld coefficient of $\gamma_0=15.0 \, \text{mJ/mol\textsubscript{La}-K\textsuperscript{2}}$ and a Debye coefficient of $\beta = 0.197 \, \text{mJ/mol\textsubscript{La}-K\textsuperscript{4}}$. More details on this fit are provided in appendix \ref{sec: Debye Fit of Heat Capacity}. The heat capacity of the $R$ = Ce compound exhibits a Sommerfeld coefficient (determined by linear extrapolation of $C(T)/T$ vs. $T^2$ in the region between $T_1^*$ and $T_2^*$) enhanced by approximately two to three times that of $R$ = La compound (see Fig. \ref{fig:A3}(b)).

\begin{figure}[H]
	\centering 
	\includegraphics[width=1\linewidth]{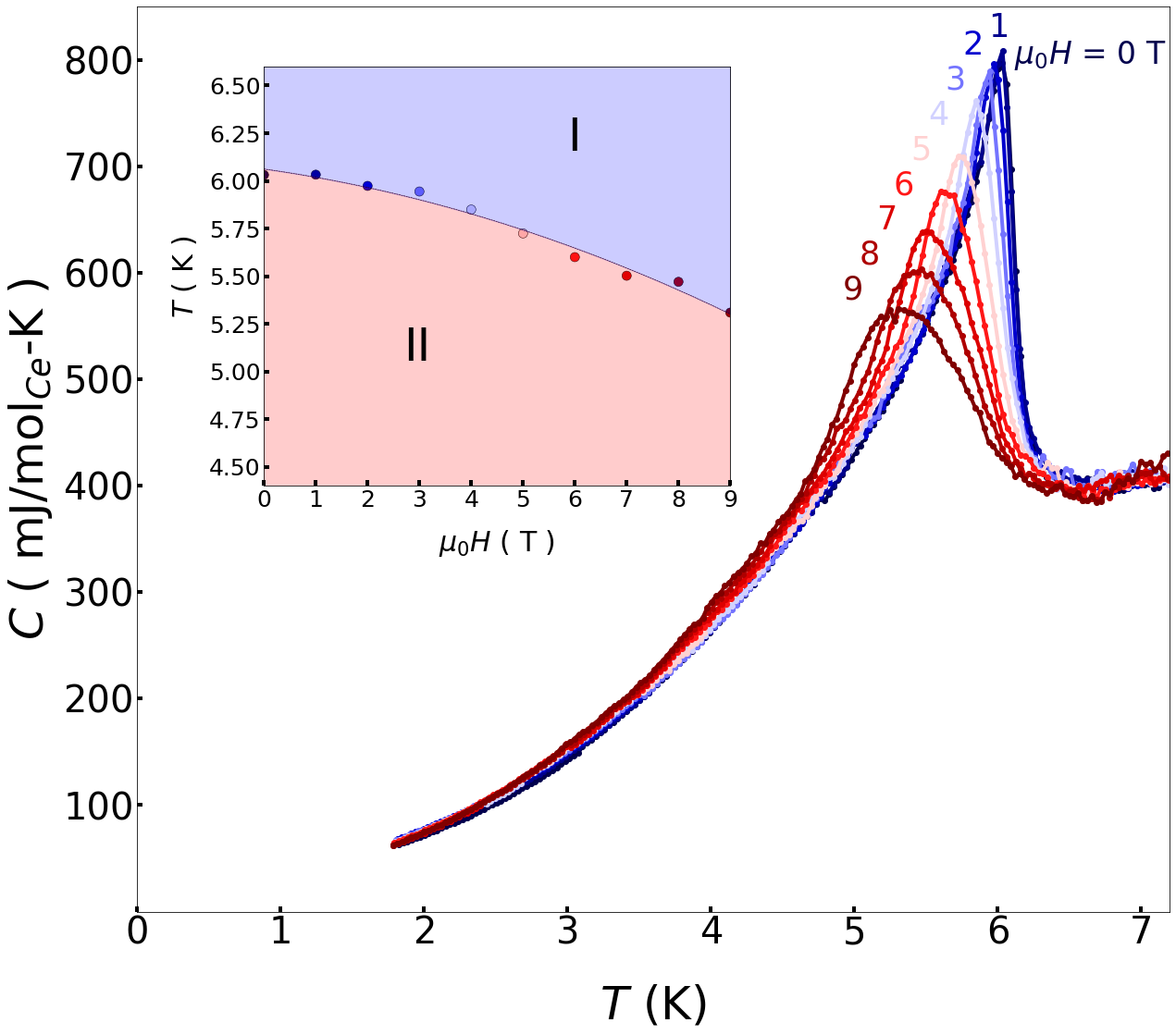}
	\caption{Ce\textsubscript{3}Ni\textsubscript{30}B\textsubscript{10} crystal \textit{C1} heat capacity plotted as a function of temperature showing the $6 \, \text{K}$ feature in increasing applied magnetic field up to $9 \, \text{T}$. The inset shows the phase space from the temperature $T_1^*$ of each peak in the main figure.}
	\label{fig:5}
\end{figure}

Figure \ref{fig:5} shows the Ce\textsubscript{3}Ni\textsubscript{30}B\textsubscript{10} heat capacity feature plotted as a function of temperature for a range of applied magnetic fields. As field is increased, the peak at $T_1^* = 6 \, \text{K}$ shifts towards lower temperatures and is broadened. The inset of Fig. \ref{fig:5} shows the phase space for Ce\textsubscript{3}Ni\textsubscript{30}B\textsubscript{10}, indicating a continuous field suppressed boundary between two regions, denoted I and II. The boundary $T^*(\mu_0 H)$ was fit by a second order polynomial to yield the zero-field slope $\frac{d T^*}{dH}|_{H=0} = -0.0383 \, \text{K/T}$. This slope may be used in conjunction with the Ehrenfest relation:

\begin{equation}
    \label{Eq: Ehrenfest} \Delta \left( \frac{\partial M}{\partial T} \right) = - \frac{d T^*}{dH} \Delta \left( \frac{C}{T} \right)
\end{equation}

\noindent
to determine the expected discontinuity in magnetization were the feature at $T_1^*$ ferromagnetic ($\Delta(X)$ indicates the discontinuous change in the quantity $X$). Crystal \textit{C1} realizes a jump of $\Delta \left(\frac{C}{T} \right) = 69.6 \, \text{mJ/mol\textsubscript{Ce}-K\textsuperscript{2}}$ (between $T_1^*$ and $T = 6.4 \, \text{K}$) which would correspond to a $\Delta \left( \frac{\partial M}{\partial T} \right) = 2.66 \, \text{emu/mol\textsubscript{Ce}-K}$. A discontinuity in the slope of $M(T)$ of this size was not observed at $T_1^*$. See appendix \ref{sec: Ce3Ni30B10 Heat Capacity Variations} for more details on this calculation.

\section{\label{sec:Discussion} Discussion}

The Ce\textsubscript{3}Ni\textsubscript{30}B\textsubscript{10} system exhibits a maximum in the heat capacity at $T_1^* = 6 \, \text{K}$ without clear corresponding features in the resistivity or magnetization. Electrically, Ce\textsubscript{3}Ni\textsubscript{30}B\textsubscript{10} behaves as a poor metal. Magnetically, the system behaves paramagnetically with a small effective moment and does not appear to order.

The longitudinal resistivity exhibits a small residual resistivity ratio ($RRR$) of approximately $1.95$ and a broad saturation with a shallow minimum at low temperatures. Similar behavior is seen in other Ce-Ni-B compounds: in CeNi\textsubscript{4}B a low temperature upturn is attributed to Kondo scattering from the fluctuating cerium valance \cite{tolinski2003mixed}, while in Ce\textsubscript{3}Ni\textsubscript{25.75}Ru\textsubscript{3.16}Al\textsubscript{4.1}B\textsubscript{10} a small $RRR$ is attributed to the presence of disorder scattering \cite{JANKA2013154}. In Ce\textsubscript{3}Ni\textsubscript{30}B\textsubscript{10}, the partial-occupancy of the nickel sites suggests that disorder may play a role in the behavior; the low temperature minimum may be connected to this.

The XPS spectrum, specifically the peak associated with the cerium $3d_{3/2}$ multiplet $3d^94f^0$ final state, indicates that Ce\textsubscript{3}Ni\textsubscript{30}B\textsubscript{10} is a mixed-valence compound with cerium valence between $\text{Ce}^{4+}$ (nonmagnetic $4f^0$) and $\text{Ce}^{3+}$ (magnetic $4f^1$). The low-temperature Curie-Weiss fit of Ce\textsubscript{3}Ni\textsubscript{30}B\textsubscript{10} results in a moment of $0.201 \, \mu_B$ per formula unit which would correspond to $0.0670 \, \mu_B$ per cerium atom were the nickel nonmagnetic (this value may then be treated as an upper bound for the effective moment per cerium site). The similarity in the low-temperature paramagnetic responses of Ce\textsubscript{3}Ni\textsubscript{30}B\textsubscript{10} and La\textsubscript{3}Ni\textsubscript{30}B\textsubscript{10} suggests that the nickel $3d$ electrons contribute in both systems. The magnetic field dependence of the magnetization seen in Fig. \ref{fig:3}(b) is also characteristic of similar mixed-valence compounds such as CeNi\textsubscript{4}B \cite{TOLINSKI2002363}. In this scenario, the small effective moment per cerium site induces a slight curve to the magnetization which may be understood as a weak Brillouin function superimposed on a relatively stronger linear Pauli magnetization from the delocalized conduction states (or possible contributions from impurities).

The anomalous heat capacity features in Ce\textsubscript{3}Ni\textsubscript{30}B\textsubscript{10} are absent in the La\textsubscript{3}Ni\textsubscript{30}B\textsubscript{10} heat capacity. The heat capacity maximum at $T_1^*$ is associated with the presence of $f$ electrons at each cerium site and does not correspond to any other features in either resistivity or magnetization. Short-ranged coherence associated with fluctuations of the $f$-states may explain the small amount of entropy associated with the $T_1^*$ feature. The mixed-valence CeNiSi\textsubscript{2} system exhibits a maximum in heat capacity at $\sim 3 \, \text{K}$ arising from the onset of short-ranged coherence of dipolar spin fluctuations; it is associated with a feature in the magnetic susceptibility, but no corresponding feature in the resistivity \cite{PhysRevB.43.10906}. A similar feature is reported in the doped Y$_{1-x}$Pr$_{x}$Ir$_{2}$Zn$_{20}$ system as PrIr\textsubscript{2}Zn\textsubscript{20} is a known antiferroquadrupolar (AFQ) material \cite{PhysRevLett.106.177001}, and the doped compound realizes a dilute system of $4f$ quadrupoles \cite{PhysRevLett.121.077206}. For a doping of $x=0.44$, the AFQ order is completely eliminated, and a weak maximum appears in the heat capacity with no corresponding sharp features in the resistivity (the feature is attributed to short-ranged correlations between the dilute quadrupoles) \cite{PhysRevLett.121.077206}.

Assuming full Ce\textsuperscript{3+} valence and the undistorted crystal field with $D_{6h}$ symmetry, the free space $^2 F_{5/2}$ ground state manifold is split into three Kramers' doublets. Although these Kramers' doublets cannot themselves host active quadrupolar order parameters due to the restrictions of time reversal symmetry, a quasi-quartet state formed from two low-lying Kramers' doublets with a small energy gap can have nontrivial quadrupolar degrees of freedom (see appendix \ref{sec: Local Crystal Field}). This analysis suggests that short-ranged quadrupolar coherence is possible in the mixed-valence system and would explain the lack of features in magnetic susceptibility and resistivity as well as the small entropy realized at $T_1^*$. The short-ranged coherence maxima seen in CeNiSi\textsubscript{2} and Y\textsubscript{1-x}Pr\textsubscript{x}Ir\textsubscript{2}Zn\textsubscript{20} (at $x=0.44$ doping) are, however, quite broad in contrast to the sharpness of the feature seen in Ce\textsubscript{3}Ni\textsubscript{30}B\textsubscript{10}. This sharpness is suggestive of a true phase transition associated with the onset of long ranged order. 

By the Ehrenfest relation, ferromagnetic ordering associated with the heat capacity at $T_1^*$ is unlikely due to the absence of any discontinuity in the slope of the magnetization with respect to temperature. Mixed-valence compounds such as CeRuSn \cite{PhysRevB.87.094421, fikavcek2013nature} and Ce\textsubscript{3}Rh\textsubscript{4}Sn\textsubscript{7} \cite{OPLETAL2022166941} which do order antiferromagnetically are associated with reduced amounts of entropy at the transition, but also show distinct features in magnetic susceptibility. A minority phase that orders antiferromagnetically may then be possible for a small volume fraction of the crystal (potentially from the nickel $3d$ electrons), though no clear corresponding features in magnetic susceptibility (or in magnetic torque, not shown here) are observed at $T_1^*$. Future neutron scattering experiments will be important to investigate any minority-phase magnetic dipole order. Scattering experiments would also be useful to probe for any structural transition at $T_1^*$, although no feature in electrical transport of the same prominence as that in heat capacity is observed.

Long-ranged ordering of electric quadrupole moments at $T_1^*$ formed from a quasi-quartet ground state is another possibility as quadrupolar ordering generally does not correspond to any feature in magnetic susceptibility (although it is associated with a smooth downturn in the resistivity as observed in PrTi\textsubscript{2}Al\textsubscript{20} ($T_Q = 2.0 \, \text{K}$) \cite{sakai2011kondo,doi:10.1143/JPSJ.81.083702}, perhaps here accounted for by a minority phase). In this scenario, below $T_1^*$ the system still realizes Curie-Weiss paramagnetic behavior associated with a small effective moment as every Kramers' doublet is expected to be magnetic; the magnetic moment may have a reduced size due to the mixed-valence nature of the $4f$-state and the high anisotropy of the effective $g$-tensor in the local $D_{6h}$ symmetric crystal field.

\section{Conclusions}
\label{sec:Conclusions}

The presented data indicates that the newly discovered mixed-valence Ce\textsubscript{3}Ni\textsubscript{30}B\textsubscript{10} system exhibits low-temperature, anomalous ordering. Notably, this system shows a peak in heat capacity at $T_1^* = 6 \, \text{K}$, which corresponds to a small loss of excess entropy per cerium site. This peak, however, does not correlate with any observable features in the magnetic susceptibility or resistivity measurements.

To determine the exact type of ordering at $T_1^*$, future research steps could include resonant ultrasound spectroscopy (RUS) and muon spin relaxation ($\mu$SR). As elastic strain couples directly to the quadrupolar moments, a characteristic softening of an elastic constant mode at $T_1^*$ probed by RUS would then provide evidence for the hypothesis of quadrupolar ordering \cite{PhysRevB.83.184434}. $\mu$SR experiments would probe whether or not the order parameter breaks time reversal symmetry by probing for a precession signal below $T_1^*$. If the system is found to break time reversal symmetry, Mössbauer spectroscopy could be used to put an upper bound on the local moment and probe any antiferromagnetic dipole ordering potentially arising from the nickel $3d$ electrons \cite{PhysRevLett.89.187202}. Resonant elastic X-ray (REX) scattering experiments could furthermore provide evidence for the realization of long-ranged order rather than short-ranged coherence below $T_1^*$ \cite{PhysRevLett.89.187202}.  Neutron scattering could be performed to probe any minority dipolar ordering.

As the anomalous ordering likely depends on the amount of nickel vacancy in the structure, future work on this system could include a nickel doping study based on a different synthesis method. Studying the behavior of this system under pressure could also shed light onto potential proximate phases. The Ce\textsubscript{3}Ni\textsubscript{30}B\textsubscript{10} system is a candidate to support a low temperature strongly correlated phase ripe for further exploration.

\begin{acknowledgments}
We acknowledge C. John for support with XPS measurements and P. Müller for support with single crystal X-ray diffraction and analysis. This work was funded, in part, by the Gordon and Betty Moore Foundation EPiQS Initiative, Grant No. GBMF9070 to J.G.C (instrumentation development) and the Army Research Office, Grant No. W911NF-16-1-0034 (material characterization).   
\end{acknowledgments}

\appendix

\section{X-ray Diffraction}
\label{sec: X-ray Diffraction}

\begin{figure}[H]
	\centering 
	\includegraphics[width=1\linewidth]{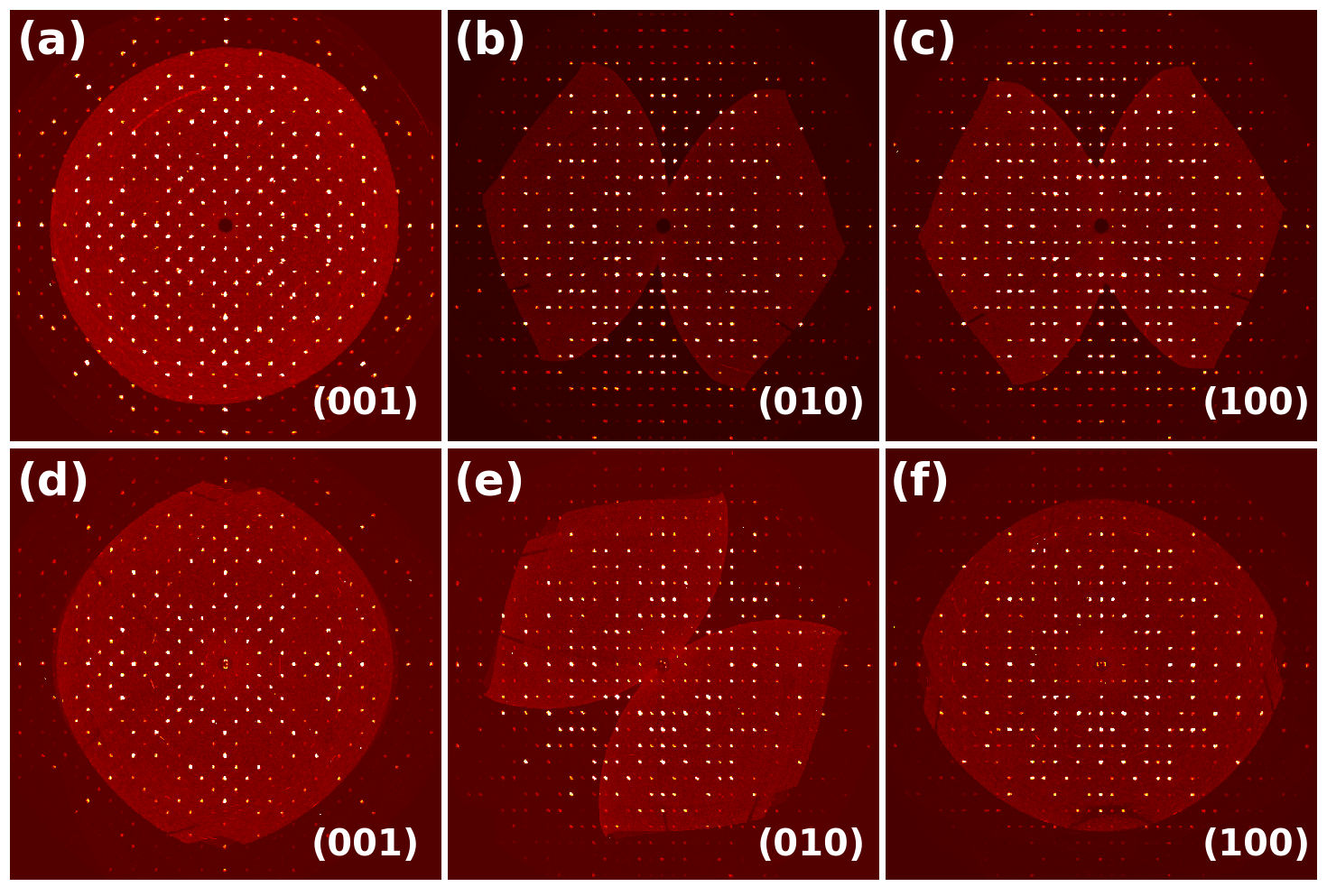}
	\caption{Single crystal X-ray diffraction patterns for the $(001)$, $(010)$, and $(100)$ planes for Ce\textsubscript{3}Ni\textsubscript{30}B\textsubscript{10} (a-c) and La\textsubscript{3}Ni\textsubscript{30}B\textsubscript{10} (d-f).}
	\label{fig:A0}
\end{figure}

Single crystal, transmission geometry X-ray diffraction was performed on both Ce\textsubscript{3}Ni\textsubscript{30}B\textsubscript{10} and La\textsubscript{3}Ni\textsubscript{30}B\textsubscript{10} at $100 \, \text{K}$. Figure \ref{fig:A0} shows the diffraction patterns obtained for the $(001)$, $(010)$, and $(100)$ planes for Ce\textsubscript{3}Ni\textsubscript{30}B\textsubscript{10} (Fig. \ref{fig:A0}a-c) and La\textsubscript{3}Ni\textsubscript{30}B\textsubscript{10} (Fig. \ref{fig:A0}d-f). Note the clear fourfold symmetry of the diffraction patterns for the $(001)$ plane.

Table \ref{T:1} shows crystallographic data and structural refinement parameters obtained using SHELXL. Table \ref{T:atomicCe} shows the Wyckoff positions, atomic coordinates, and site occupancies obtained in the refinement for Ce\textsubscript{3}Ni\textsubscript{30}B\textsubscript{10} while Table \ref{T:atomicLa} shows this data for La\textsubscript{3}Ni\textsubscript{30}B\textsubscript{10}. Note the partial occupancy for the Ni(7) and Ni(8) sites in both crystal systems.

\begin{table*}
\caption{\label{T:1}Crystallographic data and structure refinement for $R$\textsubscript{3}Ni\textsubscript{30}B\textsubscript{10} ($R$ = La, Ce), tetragonal space group $P4/nmm$}
\begin{ruledtabular}
\begin{tabular}{p{5cm} p{4cm} p{4cm}}
  Formula & La\textsubscript{3}Ni\textsubscript{30}B\textsubscript{10} &  Ce\textsubscript{3}Ni\textsubscript{30}B\textsubscript{10}   \\ [0.8ex] 
  \hline
  Lattice parameters (\AA) & $a$ = 11.3236(2) & $a$ = 11.2799(4) \\ [0.6ex] 
  $ $ & $c$ = 7.9637(2) & $c$ = 7.9354(3) \\ [0.6ex] 
  Unit cell volume (\AA\textsuperscript{3})  & 1021.14(5) & 1009.67(7) \\ [0.6ex] 
  Molar mass (g-mol\textsuperscript{-1})  & 2285.63 & 2289.26 \\ [0.6ex] 
  Crystal volume ($\mu$m\textsuperscript{3}) & 20$\times$15$\times$3 &  165$\times$90$\times$45  \\ [0.6ex]
  Calculated density (g-cm\textsuperscript{-3}) & 7.435  & 7.532  \\ [0.6ex] 
  Radiation wavelength (\AA) & 0.71073 & 0.71073 \\ [0.6ex]
  Absorption coefficient (mm\textsuperscript{-1}) & 32.976 & 33.767 \\ [0.6ex]
  F(000) & 2122 & 2128 \\ [0.6ex]
  $\theta$-range (\textsuperscript{o}) & 2.544-36.329 & 2.554-31.516 \\ [0.6ex]
  Range in $hkl$ & $\pm$18, $\pm$18, $\pm$13 & $\pm$16, $\pm$16, $\pm$11 \\ [0.6ex]
  Total reflections & 102890 & 71745 \\ [0.6ex]
  Independent reflections & 1412 ($R_{int}$ = 0.0419) & 977 ($R_{int}$ = 0.0563)\\ [0.6ex]
  Data/restraints/parameters & 1412/0/75 & 977/0/76 \\ [0.6ex]
  Goodness-of-fit on $F^2$ & 1.142 & 1.335 \\ [0.6ex]
  $R$ indices (all data) & $R_1$ = 0.0258 & $R_1$ = 0.0166, \\ [0.6ex]
  $ $ & $wR_2$ = 0.0651 & $wR_2$ = 0.0387 \\ [0.6ex]
  Largest diff. peak and hole (e-\AA\textsuperscript{-3}) & 1.479 and -1.129 & 1.376 and -1.074 \\ [0.6ex]
\end{tabular}
\end{ruledtabular}
\end{table*}

\begin{table}[H]
\caption{\label{T:atomicCe}Wyckoff positions, atomic coordinates, and site occupancies for Ce\textsubscript{3}Ni\textsubscript{30}B\textsubscript{10}.}
\begin{ruledtabular}
\begin{tabular}{p{0.8cm} P{1.1cm} P{1.7cm} P{1.7cm} P{1.7cm} P{1cm}}
  Site & Wyckoff & $x/a$ & $y/b$ & $z/c$ & Occ.\\ [0.8ex] 
  \hline
  Ce(1) & $2b$ & 0.7500 & 0.2500 & 0.5000 & 1.000\\ [0.6ex] 
  Ce(2) & $4d$ & 0.5000 & 0.5000 & 0.10000 & 1.000\\ [0.6ex] 
  Ni(1) & $8j$ & 0.3674(1) & 0.3674(1) & 0.7534(1) & 1.000\\ [0.6ex] 
  Ni(2) & $8i$ & 0.7500 & -0.0856(1) & -0.0295(1) & 1.000\\ [0.6ex] 
  Ni(3) & $8i$ & 0.7500 & 0.1395(1) & -0.1126(1) & 1.000\\ [0.6ex] 
  Ni(4) & $16k$ & 0.5742(1) & 0.0899(1) & 0.3369(1) & 1.000\\ [0.6ex] 
  Ni(5) & $8i$ & 0.4789(1) & 0.2500 & 0.5228(1) & 1.000\\ [0.6ex] 
  Ni(6) & $8j$ & 0.3607(1) & 0.1393(1) & 0.2979(1) & 1.000 \\ [0.6ex] 
  Ni(7) & $2c$ & 0.2500 & 0.2500 & 0.5336(3) & 0.435\\ [0.6ex] 
  Ni(8) & $8i$ & 0.7500 & 0.0646(1) & 0.1792(2) & 0.393\\ [0.6ex] 
  B(1) & $2c$ & 0.7500 & -0.2500 & -0.1457(11) & 1.000\\ [0.6ex] 
  B(2) & $8i$ & 0.7500 & 0.0046(4) & -0.2600(6) & 1.000\\ [0.6ex] 
  B(3) & $2c$ & 0.7500 & -0.2500 & 0.1038(11) & 1.000\\ [0.6ex] 
  B(4) & $8j$ & 0.5631(3) & -0.0631(3) & 0.4809(5) & 1.000 \\ [0.6ex] 
\end{tabular}
\end{ruledtabular}
\end{table}

\begin{table}[H]
\caption{\label{T:atomicLa}Wyckoff positions, atomic coordinates, and site occupancies for La\textsubscript{3}Ni\textsubscript{30}B\textsubscript{10}.}
\begin{ruledtabular}
\begin{tabular}{p{0.8cm} P{1.1cm} P{1.7cm} P{1.7cm} P{1.7cm} P{1cm}}
  Site & Wyckoff & $x/a$ & $y/b$ & $z/c$ & Occ.\\ [0.8ex] 
  \hline
  La(1) & $2b$ & 0.7500 & 0.2500 & 0.5000 & 1.000\\ [0.6ex] 
  La(2) & $4d$ & 0.5000 & 0.5000 & 0.10000 & 1.000\\ [0.6ex] 
  Ni(1) & $8j$ & 0.3658(1) & 0.3658(1) & 0.7495(1) & 1.000\\ [0.6ex] 
  Ni(2) & $8i$ & 0.7500 & -0.0857(1) & -0.0283(1) & 1.000\\ [0.6ex] 
  Ni(3) & $8i$ & 0.7500 & 0.1397(1) & -0.1121(1) & 1.000\\ [0.6ex] 
  Ni(4) & $16k$ & 0.5744(1) & 0.0897(1) & 0.3368(1) & 1.000\\ [0.6ex] 
  Ni(5) & $8i$ & 0.4774(1) & 0.2500 & 0.5216(1) & 1.000\\ [0.6ex] 
  Ni(6) & $8j$ & 0.3604(1) & 0.1396(1) & 0.2979(1) & 1.000\\ [0.6ex] 
  Ni(7) & $2c$ & 0.2500 & 0.2500 & 0.5299(3) & 0.409\\ [0.6ex] 
  Ni(8) & $8i$ & 0.7500 & 0.0643(1) & 0.1802(1) & 0.390\\ [0.6ex] 
  B(1) & $2c$ & 0.7500 & -0.2500 & -0.1435(10) & 1.000\\ [0.6ex] 
  B(2) & $8i$ & 0.7500 & 0.0045(3) & -0.2591(5) & 1.000\\ [0.6ex] 
  B(3) & $2c$ & 0.7500 & -0.2500 & 0.1052(10) & 1.000\\ [0.6ex] 
  B(4) & $8j$ & 0.5631(3) & -0.0620(3) & 0.4825(5) & 1.000\\ [0.6ex] 
\end{tabular}
\end{ruledtabular}
\end{table}

\section{Lanthanide Coordination Polyhedron}
\label{sec: Local Crystal Field}

\begin{figure}[H]
	\centering 
	\includegraphics[width=1\linewidth]{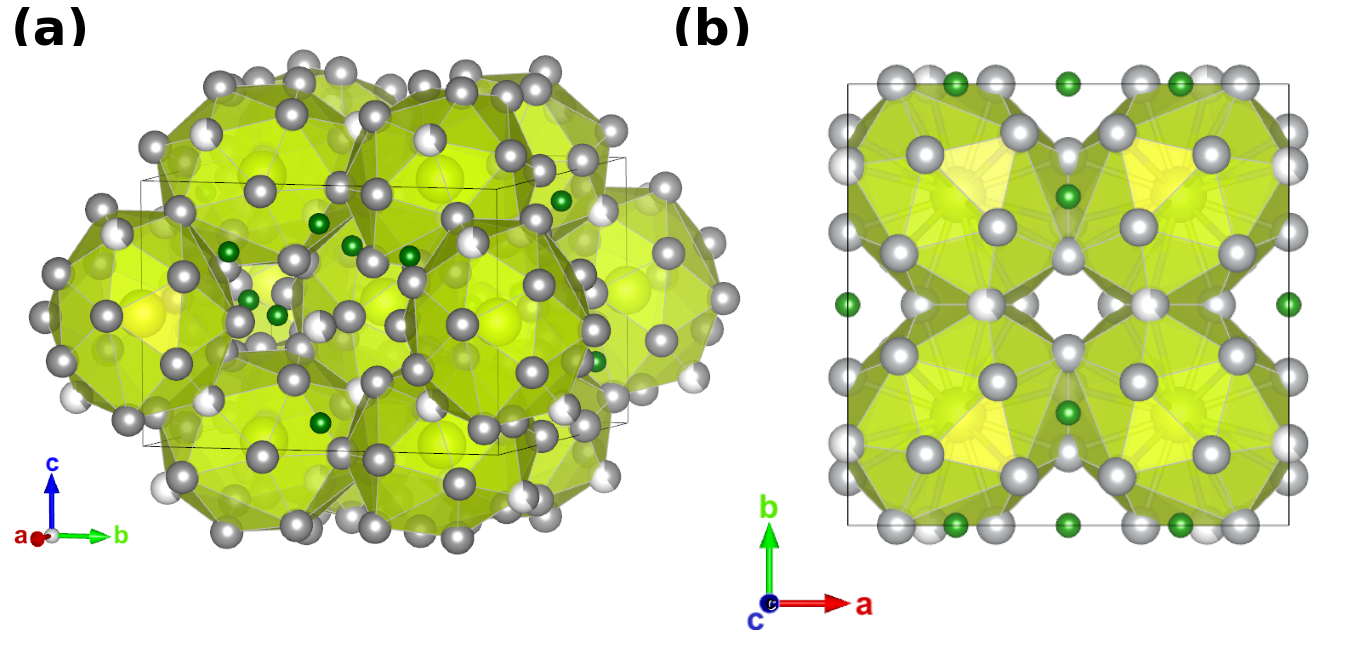}
	\caption{$R$\textsubscript{3}Ni\textsubscript{30}B\textsubscript{10} ($R$ = La, Ce) crystal structure showing the local coordination polyhedron formed by nickel sites for the lanthanide sites. (a) A view of the unit cell showing how the coordination polyhedra interlock. Note the \textit{A}-layers at the top and bottom of the unit cell with the \textit{B}-layer between them. (b) A single \textit{A}-layer of coordination polyhedra. Figure created using VESTA \cite{VESTA}.}
	\label{fig:A1}
\end{figure}

The $R$\textsubscript{3}Ni\textsubscript{30}B\textsubscript{10} ($R$ = La, Ce) structure may be understood as a structure of coordination polyhedra around each lanthanide site whereat each lanthanide atom coordinates with 20 nickel atoms. The unit cell (shown in Fig. \ref{fig:A1}(a)) is made up of two identical \textit{A}-layers of coordination polyhedra in the \textit{ab}-plane with a different \textit{B}-layer in between. The lanthanide atoms in the \textit{A}-layer (shown in Fig. \ref{fig:A1}(b)) form a square lattice with a $5.63 \, \text{\AA}$ spacing between lanthanide sites, whereas the lanthanide atoms in the \textit{B}-layer form a square lattice (rotated by 45\textsuperscript{o} relative to the \textit{A}-layers) with a $7.98 \, \text{\AA}$ spacing between lanthanide sites.

The coordination polyhedra are slightly distorted in the crystal structure such that their symmetry is lowered. Were the polyhedra undistorted, they would have a sixfold rotational symmetry axis and realize the point group $D_{6h}$ (Fig. \ref{fig:1}(c) shows a coordination polyhedron with this axis perpendicular to the page). The distortion serves to stretch each polyhedron slightly such that the symmetry is reduced from sixfold to twofold and the overall symmetry group is reduced to $D_2$.

The quadrupolar moment of the cerium $4f$ electron may be analyzed for this crystal field to check that the hypothesized quadrupolar order is possible at $T_1^*$. As the distortion in the coordination polyhedron is very slight, this analysis will be done for the undistorted polyhedron described by the point group $D_{6h}$. The Ce\textsuperscript{3+} ion will also be considered here which contains a single $4f$ electron per site. 

Due to strong spin-orbit coupling in rare earth systems, the single $4f$-electron couples to the orbital angular momentum to yield a sixfold degenerate ground state multiplet $^2 F_{5/2}$ and an eightfold degenerate excited state multiplet $^2 F_{7/2}$. Due to the large energy splitting between these multiplets, only the ground state manifold needs to be considered. The $^2 F_{5/2}$ manifold is described by $J=5/2$ angular momentum and must be analyzed using the crystal double group of the coordination polyhedron due to the odd parity of $^2 F_{5/2}$ under a $2\pi$ rotation \cite{Tinkham}. 

As the $D_{6h}$ point group is a direct product group of $D_6$ and inversion (\textit{i.e.}, $D_{6h} = D_6 \times \{E,i\}$), the CEF splitting may be analyzed using the double group $D_6'$ as the inversion operator commutes the Hamiltonian and CEF splitting cannot change the parity of the free-space eigenstates \cite{Tinkham}. The sixfold degenerate $^2 F_{5/2}$ multiplet breaks up into the three doublet irreducible representations of $D_6'$, \textit{viz.} $\Gamma_7 \oplus \Gamma_8 \oplus \Gamma_9$.

The CEF potential for a crystal field respecting $D_6$ symmetry is given by Eq. (\ref{Eq: D6}):
\begin{gather}
    \label{Eq: D6} V_{D_6} = B_0^2 C_0^2 + B_0^4 C_0^4 + B_0^6 C_0^6\\
    \label{Eq: Racah} C_q^k (\theta,\phi) = \sqrt{\frac{4\pi}{2k+1}} Y_k^q (\theta,\phi)
\end{gather}

\noindent
where $B_q^k$ are the crystal field parameters, and $C_q^k (\theta,\phi)$ are tensor operators (related to the spherical harmonics $Y_k^q (\theta,\phi)$ via Eq. (\ref{Eq: Racah})) \cite{RareEarth23}. Treating $V_{D_6}$ as a perturbation and diagonalizing it in the $\ket{J=5/2,m_J}$ basis yields the following wavefunctions:
\begin{gather}
    \label{Eq: 1/2} \Gamma_7 \quad : \quad \ket{\psi_{\pm1/2}} = \ket{5/2,\pm 1/2}\\
    \label{Eq: 3/2} \Gamma_8 \quad : \quad  \ket{\psi_{\pm3/2}} = \ket{5/2,\pm 3/2}\\
    \label{Eq: 5/2} \Gamma_9 \quad : \quad  \ket{\psi_{\pm5/2}} = \ket{5/2,\pm 5/2}
\end{gather}

\noindent
each corresponding to a Kramers' doublet in the CEF split energy level structure. Table \ref{T:Multipole} lists the irreducible representations of the $D_6'$ point group describing each of the magnetic dipole and electric quadrupole operators.

\begin{table}[H]
\caption{\label{T:Multipole}Magnetic dipole and electric quadrupole moment operators listed along with the irreducible representation of the $D_{6}'$ point group to which they belong. The form of these operators was provided by Ref. \cite{19841809}.}
\begin{ruledtabular}
\begin{tabular}{p{2cm} p{2cm} p{4cm}}
  Moment & Symmetry & Operator  \\ [0.8ex] 
  \hline
  Dipole & $\Gamma_{2}$ & $\mathbf{J}_z$ \\ [0.6ex] 
   & $\Gamma_{5}$ & $\mathbf{J}_x$ \\ [0.6ex] 
   &  & $\mathbf{J}_y$\\ [0.6ex] 
  Quadrupole & $\Gamma_{1}$ & $\mathbf{O}_2^0 = \frac13 \left( 3 \mathbf{J}_z^2 - |\mathbf{J}|^2 \right)$ \\ [0.6ex] 
   & $\Gamma_{5}$ & $\mathbf{O}_{xz} = \frac{1}{\sqrt{3}} \left( \mathbf{J}_x \mathbf{J}_z + \mathbf{J}_z \mathbf{J}_x\right)$ \\ [0.6ex] 
   &  & $\mathbf{O}_{yz} = \frac{1}{\sqrt{3}} \left( \mathbf{J}_y \mathbf{J}_z + \mathbf{J}_z \mathbf{J}_y\right) $ \\ [0.6ex] 
   & $\Gamma_{6}$ & $\mathbf{O}_2^2 = \frac{1}{\sqrt{3}} \left( \mathbf{J}_x^2 - \mathbf{J}_y^2 \right)$ \\ [0.6ex] 
   &  & $\mathbf{O}_{xy} = \frac{1}{\sqrt{3}} \left( \mathbf{J}_x \mathbf{J}_y + \mathbf{J}_y \mathbf{J}_x\right)$ \\ [0.6ex] 
\end{tabular}
\end{ruledtabular}
\end{table}

For a multipole operator $\mathbf{M}$ to become active within a Kramers' doublet spanned by $\ket{\psi_\pm}$, it must be allowed to realize a nonzero matrix element $\braket{\psi_\alpha | \mathbf{M} | \psi_\beta}$ at $T_1^*$. If a Kramers' doublet is described by the $\Gamma_w$ irreducible representation of the point group, potentially active multipoles must have an irreducible representation that is present in the decomposition of the direct product $\Gamma_w \otimes \Gamma_w$ \cite{alma990002756300106761}. As this system contains a single electron, the active multipole operators are furthermore restricted to the symmetric part of this decomposition (under particle exchange) if the operator is time reversal odd, and to the antisymmetric part if the operator is time reversal even \cite{alma990002756300106761}. Magnetic dipole operators (time reversal odd) must then appear in the symmetric part while electric quadrupole operators (time reversal even) must appear in the antisymmetric part.

The direct products for the Kramers' doublets in Eqs. (\ref{Eq: 1/2} - \ref{Eq: 5/2}) are computed within the $D_6'$ point group as follows:

\begin{gather}
    \label{Eq: 77}\Gamma_7 \otimes \Gamma_7 = \Gamma_1^A \oplus \Gamma_2^S \oplus \Gamma_5^S\\
    \label{Eq: 88}\Gamma_8 \otimes \Gamma_8 = \Gamma_1^A \oplus \Gamma_2^S \oplus \Gamma_3^S \oplus \Gamma_4^S\\
    \label{Eq: 99}\Gamma_9 \otimes \Gamma_9 = \Gamma_1^A \oplus \Gamma_2^S \oplus \Gamma_5^S
\end{gather}

\noindent
where the $A$ ($S$) superscript indicates an antisymmetric (symmetric) representation \cite{alma990002756300106761}. Although the $\Gamma_{1}$ ($\mathbf{O}_2^0$) quadrupole operator does appear in these decompositions, further computation shows that the matrix elements $\braket{\psi_\pm | \mathbf{O}_2^0 | \psi_\pm}$ are always nonzero for all the CEF split wavefunctions in Eqs. (\ref{Eq: 1/2} - \ref{Eq: 5/2})  which implies that the $\Gamma_{1}$ ($\mathbf{O}_2^0$) quadrupole moments are trivially frozen by the crystal field (without breaking time reversal symmetry) and cannot suddenly become active at $T_1^*$.

If two low-lying Kramers' doublets are separated by a small energy difference, the two states can potentially mix to form an effective quasi-quartet described by the direct sum of the irreducible representations of the two Kramers' doublets. This state could support electric quadrupole operators without violating Kramers' theorem if the irreducible representations of the relevant operators are present in the decomposition of the ``mixed'' direct product between the two Kramers' doublets. These ``mixed'' direct products are computed within the $D_6'$ point group as follows:

\begin{gather}
    \label{Eq: 78}\Gamma_7 \otimes \Gamma_8 = \Gamma_5 \oplus \Gamma_6\\
    \label{Eq: 89}\Gamma_8 \otimes \Gamma_9 = \Gamma_5 \oplus \Gamma_6\\
    \label{Eq: 79}\Gamma_7 \otimes \Gamma_9 = \Gamma_3 \oplus \Gamma_4 \oplus \Gamma_6
\end{gather}

Every direct product listed in Eqs. (\ref{Eq: 78} - \ref{Eq: 79}) contains at least one electric quadrupole representation from Table \ref{T:Multipole} which suggests that quadrupolar order at $T_1^*$ could potentially arise from a quasi-quartet formed by a pair of low lying Kramers' doublets in this system. Assuming the case of a quasi-quartet formed from degenerate $\Gamma_8$ and $\Gamma_9$ Kramers' doublets, as well as an undistorted crystal field, it can be shown that the $\Gamma_6$ ($\mathbf{O}_2^2$ or $\mathbf{O}_{xy}$) quadrupole moment does not realize any background amplitude of $\braket{\Gamma_l : \psi_\alpha | \mathbf{O} | \Gamma_{l'} : \psi_\beta}$ frozen by the crystal field (where $l,l' \in \{8, 9 \}$ and $\mathbf{O} \in \{ \mathbf{O}_2^2, \mathbf{O}_{xy}\}$). These quadrupole moment could potentially become active at $T_1^*$ for such a quasi-quartet.

\section{Curie-Weiss Fit of Magnetic Susceptibility}
\label{sec: Curie-Weiss Fit of Magnetic Susceptibility}

The magnetic susceptibility of $R$\textsubscript{3}Ni\textsubscript{30}B\textsubscript{10} ($R$ = La, Ce) was fit by a Curie-Weiss model with a constant offset via the following equation:

\begin{equation}\label{eq:X}
    \chi = \frac{1}{T-\theta_{CW}} \frac{N_A}{3k_B} \mu_{eff}^2 + \chi_0
\end{equation}

\noindent
where $\chi$ is the magnetic susceptibility per mole, $\theta_{CW}$ is the Curie temperature, $N_A$ Avogadro's number, $\chi_0$ is the constant Pauli susceptibility, and $\mu_{eff}$ is the effective moment. As both systems behave paramagnetically down to base temperature, the temperature window for the fit was chosen where $\chi_0$ was found to be approximately constant and the inverse susceptibility $(\chi-\chi_0)^{-1}$ was linear as a function of temperature.

\begin{figure}[H]
	\centering 
	\includegraphics[width=1\linewidth]{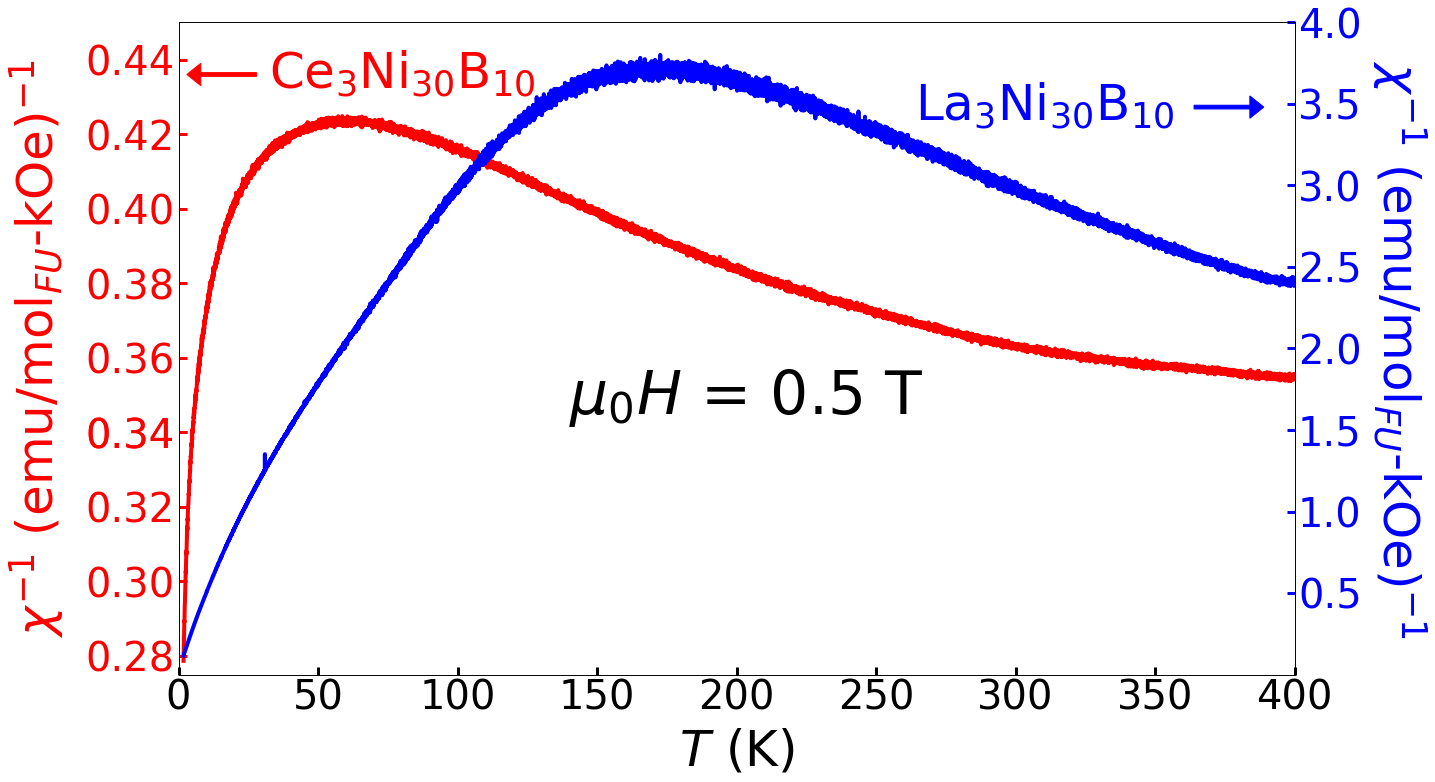}
	\caption{Inverse magnetic susceptibility data for Ce\textsubscript{3}Ni\textsubscript{30}B\textsubscript{10} crystal \textit{C2} and La\textsubscript{3}Ni\textsubscript{30}B\textsubscript{10} crystal \textit{L1} measured in $0.5 \, \text{T}$ of applied magnetic field.}
	\label{fig:A2}
\end{figure}

Figure \ref{fig:A2} shows the inverse magnetic susceptibilities for both Ce\textsubscript{3}Ni\textsubscript{30}B\textsubscript{10} and La\textsubscript{3}Ni\textsubscript{30}B\textsubscript{10}. Eq. (\ref{eq:X}) was used to fit the inverse susceptibility for Ce\textsubscript{3}Ni\textsubscript{30}B\textsubscript{10} between $1.8 - 30 \, \text{K}$ and the inverse susceptibility for La\textsubscript{3}Ni\textsubscript{30}B\textsubscript{10} between $1.8 - 50 \, \text{K}$. All the susceptibility data in Fig. \ref{fig:A2} is plotted per mole of formula unit of $R$\textsubscript{3}Ni\textsubscript{30}B\textsubscript{10} such that the effective moments of both compounds may be compared. The Curie-Weiss fit parameters for both systems are presented in Table \ref{T:2}. We note these are upper bounds considering potential contributions from impurities.

\begin{table}[H]
\caption{\label{T:2}Parameters from the Curie-Weiss fit of Ce\textsubscript{3}Ni\textsubscript{30}B\textsubscript{10} crystal \textit{C2} ($1.8 - 30 \, \text{K}$) and La\textsubscript{3}Ni\textsubscript{30}B\textsubscript{10} crystal \textit{L1} ($1.8 - 50 \, \text{K}$).}
\begin{ruledtabular}
\begin{tabular}{p{3cm} P{2cm} P{2cm}}
  Parameter &  Ce\textsubscript{3}Ni\textsubscript{30}B\textsubscript{10} &  La\textsubscript{3}Ni\textsubscript{30}B\textsubscript{10}   \\ [0.8ex] 
  \hline
  $\theta_{CW}$ (K) & -2.27 & -0.733 \\ [0.6ex] 
  $\chi_0$ (emu/mol\textsubscript{FU}-kOe) & 2.27 & 0.179 \\ [0.6ex] 
  $\mu_{eff}$ ($\mu_B$/FU) & 0.201 & 0.395 \\ [0.6ex] 
\end{tabular}
\end{ruledtabular}
\end{table}

\section{Ce\textsubscript{3}Ni\textsubscript{30}B\textsubscript{10} Heat Capacity Variations}
\label{sec: Ce3Ni30B10 Heat Capacity Variations}

\begin{figure}[H]
	\centering 
	\includegraphics[width=1\linewidth]{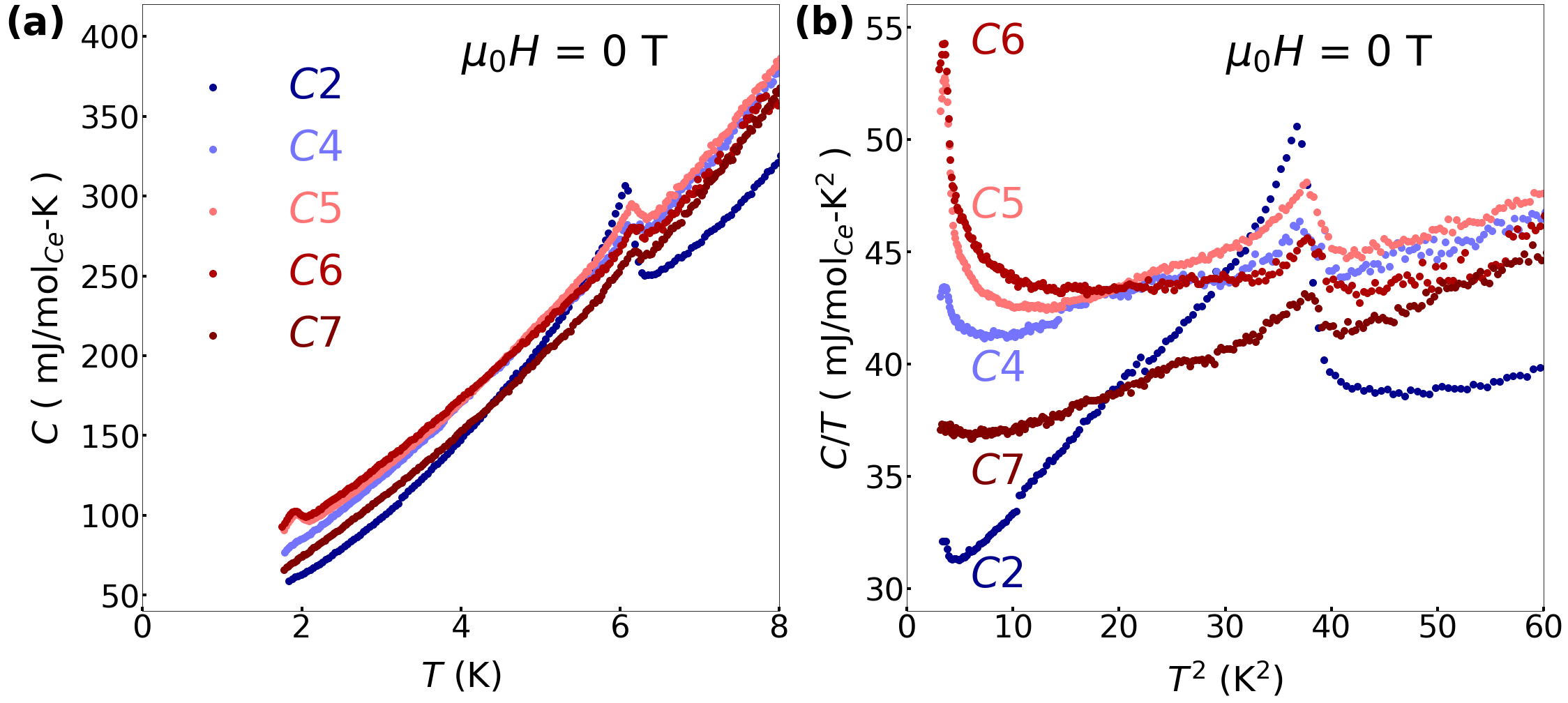}
	\caption{Comparisons of the zero-field heat capacity features at $T_1^* = 6 \, \text{K}$ and $T_2^* = 2 \, \text{K}$ for five different crystals of Ce\textsubscript{3}Ni\textsubscript{30}B\textsubscript{10} from the same synthesis batch. (a) Heat capacity plotted as a function of temperature. (b) Heat capacity divided by temperature plotted as a function of temperature squared.}
	\label{fig:A3}
\end{figure}

The features seen in the heat capacity of Ce\textsubscript{3}Ni\textsubscript{30}B\textsubscript{10} at $T_1^* = 6 \, \text{K}$ and $T_2^* = 2 \, \text{K}$ exhibit some variation in strength between samples. Figure \ref{fig:A3} shows heat capacity data for crystal \textit{C2} as well as four other crystals of Ce\textsubscript{3}Ni\textsubscript{30}B\textsubscript{10} (crystals \textit{C4}$-$\textit{C7}) from the same synthesis batch. Based on the region between $T_1^*$ and $T_2^*$ in Fig. 9(b), the Sommerfeld coefficient of Ce\textsubscript{3}Ni\textsubscript{30}B\textsubscript{10} varies from $30-45 \, \text{mJ/mol\textsubscript{Ce}-K\textsuperscript{2}}$.

\begin{table}[H]
\caption{\label{T:3}Heat capacity and entropy data associated with the $T_1^*$ feature of Ce\textsubscript{3}Ni\textsubscript{30}B\textsubscript{10} for crystals $C1$, $C2$, $C4$-$C7$. $\alpha$ is the exponent from the $C \sim T^\alpha$ fit of the background-subtracted heat capacity.}
\begin{ruledtabular}
\begin{tabular}{p{0.5cm} P{0.9cm} P{0.9cm} P{1.4cm} P{1.4cm} P{1.4cm} P{1.4cm}}
  &  $\alpha$ & $T_1^*$ & $C_{4f}^*$ & $S_{4f}^*$ & $\Delta \left(\frac{C}{T} \right)$ & $\Delta \left(\frac{\partial M}{\partial T} \right)$ \\ [0.6ex] 
  &  & K & $ \frac{\text{mJ}}{\text{mol\textsubscript{Ce}-K}} $ & $ \frac{\text{mJ}}{\text{mol\textsubscript{Ce}-K}} $ & $ \frac{\text{mJ}}{\text{mol\textsubscript{Ce}-K\textsuperscript{2}}} $ & $ \frac{\text{emu}}{\text{mol\textsubscript{Ce}-K}} $\\ [0.8ex] 
  \hline
  $C1$ & 2.50 & 6.03 & 662 & 219 & 69.6 & 2.66 \\ [0.6ex] 
  $C2$ & 1.54 & 6.06 & 171 & 94.9 & 12.3 & 0.472 \\ [0.6ex] 
  $C4$ & 0.898 & 6.08 & 145 & 156 & 3.40 & 0.130 \\ [0.6ex] 
  $C5$ & 0.866 & 6.14 & 156 & 167 & 3.55 & 0.136 \\ [0.6ex] 
  $C6$ & 0.701 & 6.14 & 141 & 191 & 2.85 & 0.109 \\ [0.6ex] 
  $C7$ & 0.942 & 6.12 & 126 & 127 & 2.49 & 0.0952 \\ [0.6ex] 
\end{tabular}
\end{ruledtabular}
\end{table}

The $4f$ electron entropy $S_{4f}^*$ associated with the $T_1^*$ feature was computed for each of the Ce\textsubscript{3}Ni\textsubscript{30}B\textsubscript{10} crystals  \textit{C1}, \textit{C2}, \textit{C4}$-$\textit{C7} and is displayed in Table \ref{T:3} along with the $4f$ electron contribution to the hear capacity $C_{4f}^*$ at $T_1^*$. The entropy was computed by first subtracting the heat capacity of La\textsubscript{3}Ni\textsubscript{30}B\textsubscript{10} to obtain the heat capacity associated with the $4f$ states of cerium and then interpolating the background-subtracted heat capacity from $2.5 \, \text{K}$ to zero via a power law fit ($C \sim T^\alpha$) of the data between $2.5 \, \text{K}$ and $5.5 \, \text{K}$. This $4f$ state heat capacity was then divided by temperature and integrated to yield the entropy associated with the $4f$ state.

\begin{figure}[H]
	\centering 
	\includegraphics[width=0.9\linewidth]{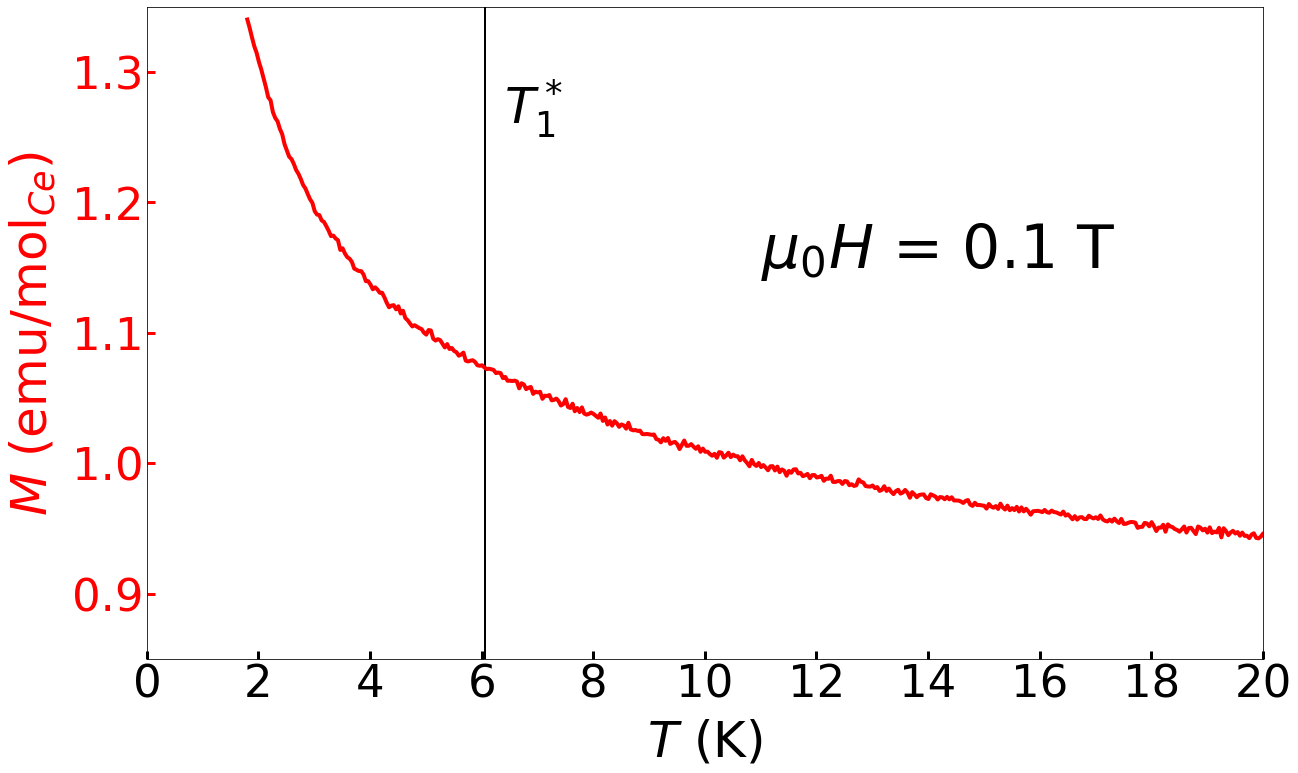}
	\caption{Magnetization data for Ce\textsubscript{3}Ni\textsubscript{30}B\textsubscript{10} crystal \textit{C2} plotted as a function of temperature in $\mu_0 H = 0.1 \, \text{T}$ of applied field. The vertical line indicates the position of $T_1^*$.}
	\label{fig:MagEhrenfest}
\end{figure}

Were the feature at $T_1^*$ ferromagnetic, the expected discontinuity in magnetization could be calculated using the Ehrenfest relation. Using the zero-field phase boundary slope $\frac{d T^*}{dH}|_{H=0} = -0.0383 \, \text{K/T}$, the expected value of $\Delta \left( \frac{\partial M}{\partial T} \right)$ was computed for all the crystals of Ce\textsubscript{3}Ni\textsubscript{30}B\textsubscript{10} listed in Table \ref{T:3} ($\Delta(\frac{C}{T})$ was computed between $T_1^*$ and $T = 6.4 \, \text{K}$). Figure \ref{fig:MagEhrenfest} shows the low temperature magnetization of Ce\textsubscript{3}Ni\textsubscript{30}B\textsubscript{10} crystal \textit{C2} with no discernible discontinuous change in slope occurring at $T_1^*$.

\section{Debye Fit of Heat Capacity}
\label{sec: Debye Fit of Heat Capacity}

\begin{figure}[H]
	\centering 
	\includegraphics[width=0.9\linewidth]{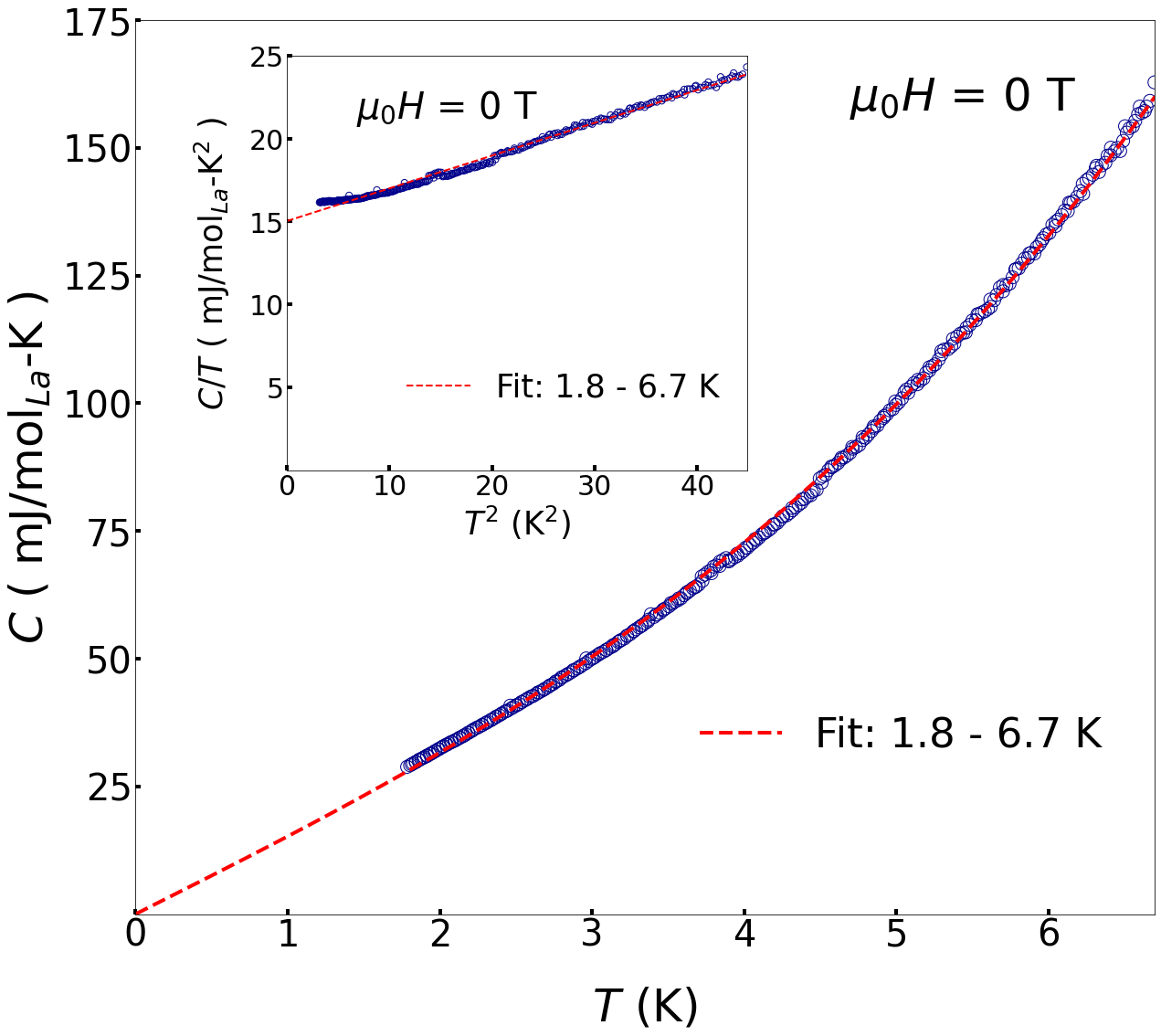}
	\caption{Zero-field heat capacity of a La\textsubscript{3}Ni\textsubscript{30}B\textsubscript{10} crystal \textit{L1} plotted as a function of temperature with a low-temperature fit of the data under $6.7 \, \text{K}$. The inset shows the data plotted as $C/T$ versus $T^2$.}
	\label{fig:A4}
\end{figure}

The heat capacity of La\textsubscript{3}Ni\textsubscript{30}B\textsubscript{10} was fit at low temperature between $1.8 \, \text{K}$ and $6.7 \, \text{K}$. The low temperature heat capacity is modeled as the sum of a linear electronic Sommerfeld term and a cubic Debye term such that:

\begin{equation}\label{eq:HC2}
    C = \gamma_0 \, T + \beta \, T^3
\end{equation}

\noindent
where $\gamma_0$ is the constant Sommerfeld coefficient, and $\beta$ is the Debye coefficient. The fit for the heat capacity of La\textsubscript{3}Ni\textsubscript{30}B\textsubscript{10} is shown in Fig. \ref{fig:A4} and results in $\gamma_0=15.0 \, \text{mJ/mol\textsubscript{La}-K\textsuperscript{2}}$ and $\beta = 0.197 \, \text{mJ/mol\textsubscript{La}-K\textsuperscript{4}}$. The inset of Fig. \ref{fig:A4} shows $C/T$ plotted as a function of $T^2$, which shows behavior that is close to linear but deviates slightly below $3 \, \text{K}$. Under the simplifying assumption that all atoms in the unit cell contribute equally to the acoustic phonon mode, this value of $\beta$ corresponds to a Debye temperature of $\theta_D = 521 \, \text{K}$ via the Debye model result:

\begin{equation}\label{beta}
    \beta = \frac{12 k_B \pi^4}{5 \theta_D^3}
\end{equation}

\noindent
where $k_B$ is the Boltzmann constant. To use Eq. (\ref{beta}), the value of $\beta$ (in units of mJ/mol\textsubscript{La}-K\textsuperscript{4}) was first divided by $14.33$ to express it as a quantity per total moles of atoms in the crystal. The experimentally determined Debye temperature for La\textsubscript{3}Ni\textsubscript{30}B\textsubscript{10} is somewhat higher than values reported for similar compounds \textit{viz.} CeNi\textsubscript{4}Al : $\theta_D = 315 \, \text{K}$ \cite{PhysRevB.70.064413}, CeNiSi\textsubscript{2} : $\theta_D = 451 \, \text{K}$ \cite{PhysRevB.43.10906}, CeNiGe\textsubscript{2} : $\theta_D = 351 \, \text{K}$ \cite{PhysRevB.43.10906}, and LaNi\textsubscript{12}B\textsubscript{6} : $\theta_D = 304 \, \text{K}$ \cite{C8DT02601G}.

\bibliography{apssamp}

\end{document}